\documentclass{ws-procs9x6}

\begin{document}
\def\vcirc{v_{\rm circ}}
\def\msun{{M_\odot}}
\def\pc{\,{\rm pc}}
\def\kpc{\,{\rm kpc}}
\def\mmsun{\,{\rm mM_{\odot}}}
\def\sech{\rm sech}
\def\kms{\,{\rm kms}^{-1}}
%
%
\def\spose#1{\hbox to 0pt{#1\hss}}
\def\lta{\mathrel{\spose{\lower 3pt\hbox{$\sim$}} \raise
2.0pt\hbox{$<$}}}
\def\gta{\mathrel{\spose{\lower 3pt\hbox{$\sim$}} \raise
2.0pt\hbox{$>$}}}

\def\Journal#1#2#3#4{{#1} {\bf #2}, #3 (#4)}
\def\MNRAS{\em MNRAS}
\def\ApJ{\em ApJ}
\def\AJ{\em AJ}
\def\Nature{\em Nature}
\def\Science{\em Science}
\def\AA{\em A\&A}
\def\AcAs{\em Acta Astron}
\def\ARAA{\em ARAA}
\def\Newa{\em New~Astron.}
\def\PhysRevLett{\em Phys.~Rev.~Lett.}

\title{RIP \\ The MACHO Era (1974-2004)}
\author{N.W. Evans, V. Belokurov}

\address{Institute of Astronomy, Madingley Rd\\
Cambridge CB3 0HA, England\\ E-mail: nwe@ast.cam.ac.uk, vasily@ast.cam.ac.uk}

\maketitle

\abstracts{This article reviews the life and death of a scientific
theory}

\section{The MACHO Ideology}

\subsection{The Dawn of Dark Matter}

The hypothesis of dark matter is often ascribed to Fritz Zwicky.
Certainly, Zwicky~\cite{fritz} in his book ``Morphological Astronomy''
noted the discrepancy between masses of clusters inferred from the
virial theorem and masses inferred from the visible constituent
galaxies. He suggests five possible explanations. The fifth (and most
tentative) --- after propositions that the clusters may not be in
equilibrium or that light may tire on traversal of enormous distances
--- is: ``{\it Finally, attention must be called to the recent
discovery of luminous and of dark intergalactic matter. The existence
of this dark matter may seriously affect all previous estimates
concerning the distribution of mass in the Universe}''.

The focus of this conference is on the direct and indirect detection
of dark matter in the Milky Way galaxy and other nearby galaxies. Even
if Zwicky was the first to hypothesise the existence of dark matter in
clusters, he did not believe that there was appreciable dark matter in
galaxies (in ``Morphological Astronomy'', he advocated Keplerian
fitting to rotation curves to estimate the masses of galaxies). The
realisation that galaxies are surrounded by dark matter haloes only
came much later. Dark matter on the scales of galaxies became widely
accepted after the publication of the rotation curve of the nearby
galaxies M31, M81 and M101 by Roberts and
collaborators~\cite{roberts}. In an influential paper, Ostriker,
Peebles \& Yahil~\cite{opy} brought together a number of lines of
evidence to suggest that: ``{\it There are reasons, increasing in
number and in quality, to believe that the masses of ordinary galaxies
may have been underestimated by a factor of 10 or more\dots The very
large implied mass to light ratios and very great extent of spiral
galaxies can perhaps most plausibly be understood as due to a giant
halo of faint stars}''

This is the first statement of the MACHO ideology -- namely that (some
of) the dark matter in galaxy haloes is baryonic and composed of
massive objects. The most obvious candidates are faint stars (red
dwarfs, white dwarfs, neutron stars), failed stars (brown dwarfs and
Jupiters) and massive remnants from an early epoch of Population III
stars. The neologism MACHO seems to have been first used in print by
Griest~\cite{kim} as a witty counterpoise to WIMPS (weakly interacting
massive particles). MACHO stands for massive compact halo objects.
 
\subsection{The Hey-Day of the MACHO Era (1974-1994)}

The Zeitgeist is well documented in the Princeton conference on ``Dark
Matter in the Universe'', which marks the hey-day of the MACHO Era.
It was well-known that all the dark matter in galaxies and clusters
could conceivably be baryonic without violating constraints from
cosmological nucleosynthesis~\cite{martin}. There even seemed to be
arguments in favour of baryonic compact objects as opposed to particle
dark matter. For example, Gunn~\cite{jim} pointed out that; ``{\it
There is evidence that the Population II mass function is very steep
in the halo and an extension at the low mass-end to quite plausible
masses leads to very large mass-to-light ratios\dots A picture in
which the low-mass cut-off progresses smoothly from 0.1 $\msun$ to
$10^{-3}$ $\msun$ as one goes from the center of the galaxy outwards
makes a qualitatively plausible model\dots It entails no mystery as to
why the amount of dark matter is within an order of magnitude of the
visible matter, and makes plausible the fact that rotation curves are
flattish from regions where the galaxies are dominated by visible matter
out to regions in which they are dominated by dark matter}.''

More exuberantly still, Lynden-Bell~\cite{dlb} cited the X-ray data;
``{\it We have rather good evidence that around a number of giant
elliptical galaxies, baryonic matter is disappearing from hot, X-ray
emitting gas. The place where it disappears is right for the making of
dark halos. The rate of its disappearance would build a halo in
$10^{10}$ years. If we want to believe the observations rather than
our prejudices, we should take as our best bet that dark halos are
baryonic and made from cooling flows\dots When exotic neutral particles
have been found in the laboratory, I shall be happy to postulate them
in the cosmos, but until then, let us use our observations, not our
prejudices.}''
 
But even then, the most important objection to baryonic dark matter as
the dominant component of galaxy haloes was clearly understood. It is
difficult to understand how such baryonic structures of mass $\sim 10^{12}
\msun$ could have formed without leaving an imprint in the microwave
background~\cite{jim}.

\subsection{The Decline and Fall of the MACHO Era (1994-2004)}

Microlensing as a test for dark, compact objects was suggested very
early on (e.g., Zwicky's ``Morphological Astronomy'' discusses
microlensing by neutron stars).  But, Paczy\'nski~\cite{bohdan}
convinced the astronomical community that microlensing could provide a
decisive test of the MACHO hypothesis. And so it turned out ....  The
microlensing experiments led to the decline and fall of the MACHO Era.

Beginning in 1993, large scale monitoring of stars in the Large
Magellanic Cloud (LMC) was conducted by two groups (MACHO and EROS)
looking for microlensing events.  The results of the MACHO experiment
are well-known.  From 5.7 years of data, Alcock et al.~\cite{alcock}
found between 13 to 17 microlensing events and reckoned that the
microlensing optical depth (or probability of microlensing) is $\tau
\sim 1.2^{+0.4}_{-0.3} \times 10^{-7}$. Interpreted as a dark halo
population, the most likely fraction of the dark halo in MACHOs is 20
\%, while the most likely mass of the MACHOs is between 0.15 and 0.9
$\msun$.  After 8 years of monitoring the Magellanic Clouds, the EROS
experiment announced $3$ microlensing candidates towards the
LMC~\cite{thierry}. Although EROS do not report their results in terms
of optical depth, they have clearly detected a smaller microlensing
signal than MACHO -- a discrepancy which could have a number of
explanations.

The remainder of this article will argue that Alcock et al.
overestimated the microlensing optical depth and that the dark halo
has little or no MACHOs.

\begin{figure}[t]
\centerline{\psfig{figure=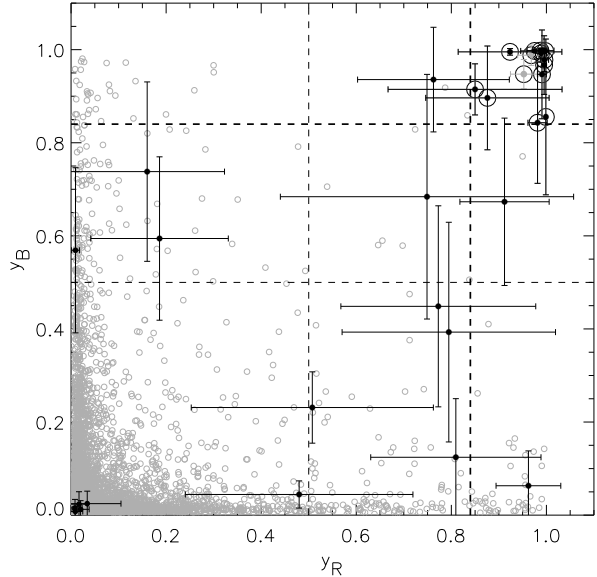,height=8.cm}}
\caption{The locations of $\approx 22000$ lightcurves as given by the
outputs of the neural networks $y_R$ and $y_B$ on processing the red
data and the blue data respectively. These include the 29 lightcurves
that passed the loose selection of the MACHO collaboration together
with $\sim 1000$ lightcurves in the vicinity of each candidate. Each
point gives the maximum of the moderated output while the error bar
gives the network scatter. A large open circles around a point
indicates that it lies above the decision boundary ($y_R > 0.87$ and
$y_B > 0.87$). Filled black dots represent the 29 lightcurves selected
by Alcock et al., while all other lightcurves are represented by open
grey dots. [From Belokurov et al. 2004]}
\label{fig:manytiles}
\end{figure}
\begin{figure}[t]
\centerline{\psfig{figure=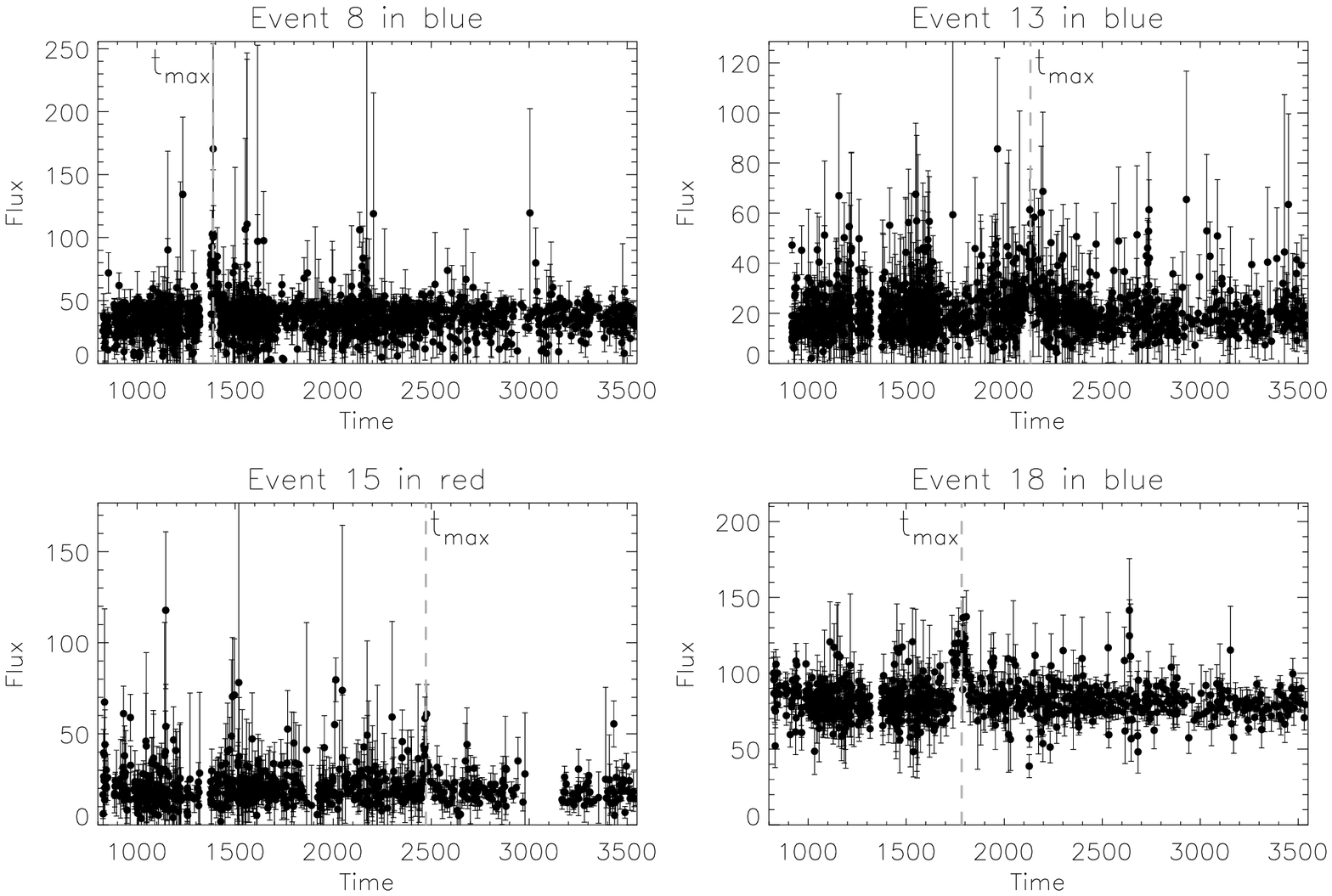,height=7.cm}}
\caption{This shows the lightcurves for 4 events which received low
probability values $y$ in one or both filters. These are all included
in Alcock et al.'s (2000) set of convincing microlensing
candidates, but are not confirmed by our neural network analysis.  The
vertical axis is flux in ADU s$^{-1}$ and the horizontal axis is time
in JD-2448000. Vertical lines mark the peak of the event. [From
Belokurov et al. 2004]}
\label{fig:miss}
\end{figure}

\section{Neural Network Processing}

There are two principal difficulties with the microlensing
experiments.  The first is well-known, the second less so (and thus we
concentrate upon it here).

First, just as in direct detection experiments for particle dark
matter, there is a background that must be eliminated. In microlensing
experiments, stars in the thin disk, thick disk and the LMC all provide
possible lenses for microlensing events~\cite{noofs}, aside from
MACHOs in the dark halo. The total optical depth due to stellar
lensing from known populations~\cite{opds} is $\sim 0.7 \times
10^{-7}$, which is within the $2 \sigma$ lower bound of Alcock et
al.'s claimed detection ($\tau \sim 1.2^{+0.4}_{-0.3} \times
10^{-7}$).
 
Secondly, the identification of microlensing events (stars that
brighten and then fade) takes place against a background of stellar
variability that is at least $10^5$ times more common. Many varieties
of stellar variability are not well-studied or understood. Therefore,
the identification of events is much more fraught than usually
appreciated. All microlensing groups use a sequence of straight line
cuts to identify events (for example, excising chromatic lightcurves
or troublesome regions of the colour-magnitude diagran). The decision
boundary between microlensing and non-microlensing is therefore
polygonal in a multi-dimensional parameter space.  Nowadays, many
high-energy physics experiments prefer to use neural networks for
pattern recognition. This is because neural networks permit the
construction of complicated decision boundaries.

All this inspired Belokurov, Evans \& Le Du~\cite{bel} to carry out a
re-analysis of the MACHO data with neural networks. Microlensing
events are characterised by the presence of (i) an excursion from the
baseline that is (ii) positive, (iii) symmetric and (iv) single.  The
event itself is parameterised by (v) a timescale.  Motivated by these
features, five parameters are extracted from the lightcurves as inputs
to the neural networks.  Most neural networks require a training set,
on which the network is taught to recognise the desired patterns (in
this case, microlensing). Here, the training set contains 1500
examples of microlensing and $>$ 2000 examples of other kinds of
variability (pre-main sequence stars, Coronae Borealis stars, Miras,
Semi-regular variables, Cepheids, Bumpers, Supernovae, novae,
eclipsing variables). They are sampled with MACHO sampling and random
Gaussian noise is added.  All networks are trained using the Netlab
package~\cite{nabney}. The output of the network is the posterior
probability of microlensing.

Figure~\ref{fig:manytiles} shows the locations of $\approx 22000$
lightcurves.  The data for the red and blue passbands are processed
separately with neural networks to give outputs $y_R$ and $y_B$. The
decision boundary is shown in the bold broken line -- convincing
microlensing candidates have $y_{R,B} > 0.84$.  This boundary is fixed
by insisting that the number of false negatives in the entire MACHO
dataset is $\lta 1$. This corresponds to a false positive rate of $0.3
\%$. The 29 candidate microlensing lightcurves identified by Alcock et
al.~\cite{alcock} are denoted by filled black dots, while all other
lightcurves are shown as open grey dots. Twelve of these 29
lightcurves satisfy $y_{R,B} > 0.84$, namely 1a, 1b, 5, 6, 10a, 11,
14, 21-25. There are additionally 2 false positives. Both lie close to
the noise/microlensing border in parameter space.

After successfully passing the first tier of neural networks,
Belokurov et al.~\cite{bel} apply a second tier that discriminates
against supernovae (SNe) occurring in background galaxies behind the LMC.
The colours change dramatically during a supernova explosion as a
result of complicated radiation processes inside the ejecta. After a
fairly constant pre-maximum epoch with $B-V \approx 0$, a supernova of
type Ia typically starts turning red at the time of the maximum light.
It reaches $B-V \approx 1$ in about 30 days and then drops
back~\cite{phillips}. This can be contrasted with the colour behaviour
during gravitational microlensing. Gravity bends light irrespective of
its frequency. Therefore, colour does not change during
microlensing. However, the achromaticity of the lightcurve only holds
good if the source star is resolved and the lens is dark. The presence
of other stars within the centroid of light or lensing by a luminous
object will result in a colour change during the event. At the
baseline, the colour is defined by the combined flux from all the
sources. The amplified star will contribute most of the colour around
the peak. The colour of a microlensing event can become redder or
bluer, depending on the population of the blend, but it usually
changes symmetrically about the peak with substantial correlation
between passbands~\cite{rosanne}. The differing behaviour of colour
evolution during SNe and blended microlensing can be quantified as
features fed to neural networks, and -- as Belokurov et al.~\cite{bel}
show -- used to distinguish between the two. This leads to the
discarding of a further 3 of the 12 candidates that passed the first
tier.
                                                                           
Based on a neural network analysis, Alcock et al.'s sample is
seriously contaminated. There are 6 almost certain microlensing events
(1, 5, 6, 14, 21 and 25) and two likely ones (9, and 18).  Some of the
lightcurves rejected by the neural networks, but classified as
microlensing by Alcock et al., are shown in Figure~\ref{fig:miss}.
The peak of the alleged event is shown as a vertical dashed line.
Notice that event 23 -- which looks perfect and passes all the neural
networks -- has been shown by the EROS collaboration~\cite{eros} to
have a second peak on the lightcurve about 7 years after the first one
probed by MACHO and so is an unusual variable star. This is very
worrisome for all microlensing experiments.

\begin{figure}[t]
\centerline{\psfig{figure=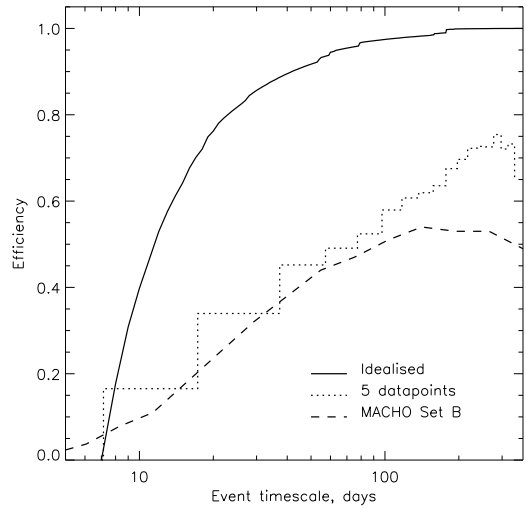,height=7.cm}}
\caption{This shows three approximations to the efficiency as a
function of Einstein diameter crossing timescale $t$.  The full
curve is the idealised efficiency calculated by integration over the
distribution of sampling gaps. This is an upper limit. The dashed
curve is the published efficiency of Alcock et al. (2000). This is a
lower limit.  The histogram is a realisation of the true efficiency
derived from Monte Carlo simulations of event selection in the
training set.}
\label{fig:epsilon}
\end{figure}

\section{The Optical Depth}

The conventional formula for optical depth is
\begin{equation}
\tau = {\pi \over 4} {1 \over N_\star T} \sum_j {t_j \over
\epsilon(t_j)}
\label{eq:estimator}
\end{equation}
where $t_j$ is the Einstein diameter crossing time of the $j$th event,
$\epsilon$ is the efficiency, $T$ is the duration of the experiment
and $N_\star$ is the number of stars monitored. The summation is taken
over the set of microlensing events. There are three major components
to the efficiency. The first arises from shortcomings in the cuts used
to identify microlensing events. The second arises from blending,
which causes both the magnification and the number of stars monitored
to be underestimated.  The third arises from the temporal sampling, as
events are necessarily missed if they fall in a gap in the
data-taking.  A neural network, properly trained, will all but
eliminate any contribution from the first component for the subset of
events included in the training set. The second component cancels out
to lowest order, as the loss due to the underestimate of the
magnification is balanced by the gain due to the fact that an object
may contain more than one star~\cite{afonso}. The third component of
the efficiency still remains, but fortunately is straightforward to
compute.
                                                                               
Figure~\ref{fig:epsilon} shows upper and lower bounds to the
efficiency as a function of timescale. An upper bound to the
efficiency can be found by assuming that events are missed if and only
if no data are taken during the bump. We sum the distribution of
sampling gaps over the baseline of the experiment and judge an event
to be missed if it falls within a gap. The probability of missing an
event with timescale $t$ is just
\begin{equation}
P(t) \propto \sum_{t' \ge t} t'n(t')
\end{equation}
where $n(t)$ is the number of gaps of duration $t$. The quantity
$1-P(t)$ is an idealised efficiency which is shown as a full curve in
Figure~\ref{fig:epsilon}. A lower bound to the efficiency is given by
the published efficiency results of Alcock et al.~\cite{alcock} for the
looser set of candidates.  This is because the neural networks
necessarily provide a cleaner set of microlensing candidates,
uncontaminated by spurious events. A realisation of the actual
efficiency is easily found from Monte Carlo simulations of the
training set, by finding the fraction of all events that are included
(and hence will be inexorably characterised as microlensing by the
network). In the simulations, microlensing events are generated with
uniform priors. Only those events with five or more datapoints with a
signal-to-noise greater than 5 are incorporated into the training
set. The efficiency is therefore the ratio of events accepted into the
training set to all events. The result is shown as a histogram in
Figure~\ref{fig:epsilon}, and lies between the upper and lower bounds,
as expected.
 
Applying eq.~(\ref{eq:estimator}) to the set of 9 events found by the
neural network, we obtain the following bound on the optical depth to
the LMC:
\begin{equation}
3 \times 10^{-8} < \tau < 5 \times 10^{-8}.
\end{equation}
Here, the timescales uncorrected for blending given in third column of
Table 7 of Alcock et al.~\cite{alcock} are used. This is correct, as
the effects of blending cancel out to lowest order.

This is a low value for the optical depth. The optical depths of the
thin disk, thick disk and spheroid to be $2.2 \times 10^{-8}$, whilst
the optical depth of the stellar content of the LMC to be $3.2 \times
10^{-8}$ on average. In other words, {\it our total optical depth
matches the contribution from the known stellar populations in the
outer Galaxy and the LMC. This implies that there is no contribution
needed from compact objects in the halo.}

There is supporting evidence for this belief from the exotic events
and from the lensing signal towards the Small Magellanic Cloud (SMC).
First, the exotic events yield additional information which can break
some of the microlensing degeneracies and thus give indirect evidence
on the location of the lens. All the exotic lenses belong to known
stellar populations in the outer Milky Way or the LMC.  Second, the
duration of the events towards the SMC is very different from the
duration towards the LMC. The EROS collaboration~\cite{afonsoag}
constrain the optical depth towards the SMC to be $< 10^{-7}$ at
better than the 90 \% confidence level, based on an admittedly small
sample. Both these facts militate against the idea that a single
population of objects in the Milky Way halo is causing the
microlensing events

\section{Conclusions}

The MACHO Era is over! The dark matter in the halo of the Milky Way is
{\bf not} in the form of massive, compact halo objects. The
microlensing signal detected by both the MACHO and EROS experiments is
entirely consistent with that expected from stellar lenses in the
known populations. In particular, the sample of 14 high quality
microlensing events in Alcock et al.~\cite{alcock} is
contaminated. Realistically, Alcock et al.'s sample has 6 almost
certain microlensing events (1, 5, 6, 14, 21 and 25) and two likely
ones (9, and 18). This is consistent with expectations from known
stellar populations.

Even for the die-hards, the matter will surely soon be settled by the
POINT-AGAPE experiment~\cite{pa}. This is a microlensing experiment
towards the nearby Andromeda galaxy (M31), which probes a new line of
sight through the Milky Way and M31 dark haloes. It will provide a new
estimate of the fraction of the Milky Way and M31 dark haloes that is
composed of MACHOs.  Two fields north and south of the M31 bulge have
been monitored for three years using the Wide Field Camera on the
Isaac Newton Telescope. The POINT-AGAPE collaboration have already
found a small number of interesting individual microlensing events
towards M31~\cite{pa}, carried out a survey for classical
novae~\cite{matt} and reported the locations, periods and brightness
of $\sim 35000$ variable stars~\cite{jin}. Very recently, an
unrestricted and fully automated search for microlensing events
towards M31 has been published~\cite{belpa}. Using a series of seven
cuts based on sampling, goodness of fit, consistency, achromaticity,
position in the colour-magnitude diagram and signal-to-noise. This
leaves just 3 first-level or convincing microlensing candidates and 3
second-level or possible microlensing candidates. The efficiency of
this survey is being computed at the moment and will yield an
independent estimate of the MACHO fraction.

Die-hards have only a short time to wait.


\begin{thebibliography}{99}


\bibitem{fritz}F. Zwicky, {\it Morphological Astronomy},
  Springer-Verlag, Berlin (1957)

\bibitem{roberts} M.S. Roberts, A.H. Rots, \Journal{\AA}{26}{483}{1974};
M.S. Roberts, R.N. Whitehurst, \Journal{\ApJ}{201}{327}{1975}

\bibitem{opy} J.P. Ostriker, P.J.E. Peebles, A. Yahil,
\Journal{\ApJ}{193}{L1}{1974}

\bibitem{kim} K. Griest, \Journal{\ApJ}{366}{412}{1991}

\bibitem{martin} M.J. Rees, 
In {\it IAU Symposium 127: Dark Matter in
  the Universe}, eds. J. Kormendy, G.R. Knapp, Reidel, Dordrecht, p. 396

\bibitem{jim} J.E. Gunn, In {\it IAU Symposium 127: Dark Matter in
  the Universe}, eds. J. Kormendy, G.R. Knapp, Reidel, Dordrecht, p. 543 

\bibitem{dlb} D. Lynden-Bell, In {\it IAU Symposium 127: Dark Matter in
  the Universe}, eds. J. Kormendy, G.R. Knapp, Reidel, Dordrecht,
  p. 530 

\bibitem{bohdan} B. Paczy\'nski, \Journal{\ApJ}{304}{1}{1986}

\bibitem{alcock} C. Alcock et al., \Journal{\ApJ}{542}{281}{2000}
 
\bibitem{thierry} T. Lasserre et al., \Journal{\AA}{355}{L39}{2000}
 
\bibitem{noofs} K. Sahu, \Journal{\Nature}{370}{275}{1994};
N.W. Evans, G. Gyuk, M.S. Turner, J.J. Binney,
\Journal{\ApJ}{501}{L45}{1998}; H.S. Zhao,
\Journal{\MNRAS}{294}{139}{1998}; H.S. Zhao, N.W. Evans,
\Journal{\ApJ}{545}{L35}{2000}

\bibitem{opds} C. Alcock et al., \Journal{\ApJ}{479}{119}{1997}

\bibitem{bel} V. Belokurov, N,W. Evans, Y. Le Du, \Journal{\MNRAS}{352}{233}{2004}

\bibitem{nabney} I.T. Nabney, {\it  Netlab}, Springer-Verlag, New York
(2002)

\bibitem{phillips} M.M Phillips, P. Lira, N.B. Suntzef, R.A. Schommer, 
M. Hamuy, J. Maza, \Journal{\AJ}{118}{1766}{1999}

\bibitem{rosanne} R. Di Stefano, A.A. Esin, \Journal{\ApJ}{448}{L1}{1995}

\bibitem{eros} J.F. Glicenstein, Talk at the Hawaiian 
Gravitational Microlensing Workshop, 2004
 
\bibitem{afonso} C. Afonso et al., \Journal{\AA}{400}{951}{2003} 

\bibitem{afonsoag} C. Afonso et al., \Journal{\AA}{404}{145}{2003} 

\bibitem{pa} S. Paulin-Henriksson et al., \Journal{\AA}{405}{15}{2003};
S. Paulin-Henriksson et al.,\Journal{\ApJ}{576}{L121}{2002}; J. An et al.,
\Journal{\ApJ}{601}{845}{2004}

\bibitem{matt} M. Darnley et al., \Journal{\MNRAS}{353}{571}{2004}

\bibitem{jin} J. An et al., \Journal{\MNRAS}{351}{1071}{2004}

\bibitem{belpa} V. Belokurov et  al., MNRAS, in press (astro-ph/0411186)

\end{thebibliography}
\end{document}